\documentclass{article}
\usepackage[utf8]{inputenc}
\usepackage{multirow}
\usepackage{multicol,lipsum}
\usepackage{amsmath}
\usepackage{graphicx}
\usepackage{caption}
\usepackage{subcaption}
\usepackage{authblk}
\title{Interacting Dark Energy: New parametrization and observational constraints}
\author{Arkajit Aich}
\affil{Department of Basic Sciences, Atria University, \\ 1st Main Rd, Chola Nagar, Anandnagar, Hebbal, Bengaluru - 560024, Karnataka, India}
\date{} %
\begin{document}

\maketitle

\begin{abstract}
We have re-investigated Cosmology involving interaction between Dark matter and Dark Energy in the light of a new parametrization. The new parametrization is based on the hypothesis that when Dark matter and Dark Energy will interact, Dark matter will dilute in a different manner than standard non-interacting scenario. We re-built the Cosmological equations with this new parametrization. Observational constraints on the traditional Cosmological parameters and new parameters has also been obtained by using supernova data from Pantheon and Hubble data. The parameter values obtained are $H_0$ = 69.023 $\pm$ 0.722, $M$ = -19.385 $\pm$ 0.019, $I$ = 2.901 $\pm$ 0.092 and $\Omega_{dm0}(1 - \frac{3}{I})\frac{1}{\kappa}$ = 0.254 $\pm$ 0.023 where $H_0$, $M$, $\Omega_{dm0}$ and $\kappa$ are Hubble constant, absolute magnitude of type 1a supernova, present day dark matter density and coupling parameter between dark matter and dark energy respectively while $I$ is a new parameter, dubbed dilution parameter which we introduced in the model representing the modified dilution of dark matter in the interacting scenario. The physical features of the model in regard to evolution of the Universe, deceleration parameter, age of Universe, particle physics implications of interacting scenario has also been explored in depth and conclusions has been drawn.       
\end{abstract}

\section{Introduction}

The advent of Einstein's celebrated theory of General Relativity(GR)\cite{Einstein1} in 1915, paved the pathway for a new era in Cosmology due to the development of numerous GR-based Cosmological models in the following decades ranging from static Universe\cite{Einstein2} to expanding Universe\cite{Lemaitre}. Furthermore, it was soon revealed that the Universe, besides having ordinary matter and radiation also harbours two exotic components viz. Dark matter and Dark Energy. The concept of Dark matter was first introduced in 1933 when Zwicky inferred the presence of unseen matter in the Universe from studying Coma cluster \cite{Zwicky}. The major differentiating feature of Dark matter from ordinary baryonic matter is that they do not exhibit electromagnetic, strong or weak interactions like baryonic matter but does exhibit similar gravitational interaction. Dark Energy entered the picture much later in 1998 when the international collaborations of Supernova Cosmology project \cite{Riess} and high-z supernova search team \cite{Perlmutter} inferred from analysing supernova data that the expansion rate of the Universe is accelerating and the existence of an exotic physical component in the Universe with negative pressure, namely, Dark Energy, was required to explain the observed acceleration within the framework of General Relativity. The two unknown components i.e. Dark Matter and Dark Energy together makes up about 96\% of the Universe \cite{WMAP}. Understanding the dark sector of the Universe is one central goals of modern Cosmology. In standard Cosmology, it is usually assumed  that the two components of the dark sector do not interact with each other. However, there is no physical basis to assume so and there is always a possibility that the two components interact non-minimally with each other. This has motivated a large number of studies where Cosmological scenario involving interaction between Dark Energy and Dark matter has been considered\footnote{Interested readers may see the review\cite{Wang1} for an account of various works done on Cosmological models involving interaction in the dark sector} \cite{Amendola1, Mangano, Bolotin, Costa1, Khurshudyan, Biswas1, Biswas2, Boehmer, Tamanini, Chen, Harko, Chimento1, Chimento2, Wang2, Pan1, Biswas3, Amendola2, Farrar, Pettorino1, Pettorino2, Gavela, Cabral, Pourtsidou, Bonometto1, Nunes1, Faraoni, Amendola3, Skordis, Maccio, Bonometto2, Casas, Sharov1, Bonometto3}. Such models often regarded as Interacting Dark Energy models (IDE) are compatible with observations \cite{Yang1, Salvatelli1, Nunes2, Amendola4, Amendola5, Amendola6, Manini, Vergani, Abdalla, Honorez, Baldi1, Baldi2, Beynon, Pettorino3, Salvatelli2, Piloyan1, Piloyan2, Yang2, Ferreira, Murgia, Pan2, Feng, Santos, Wang3, Marttens, Costa2, Bachega} as well. In Cosmology, IDE scenario is mainly introduced through some unknown interaction term that appears in the energy conservation equations of dark matter and dark energy, where the expression of unknown interaction term is usually chosen in a phenomenological manner. This results in a system of equations and the conventional prescription to handle such a situation is to perform a dynamical system analysis. However, the complexity of dynamical system approach can be completely side-stepped if some additional assumptions is introduced in the system. In this work, we put forward the assumption that in IDE Cosmology, since Dark matter is interacting with Dark Energy, it will dilute in a manner different from $\Lambda$CDM Cosmology. Based on this hypothesis, it is possible to obtain a much more simpler yet effective parametrization of IDE scenario which we will show in the following sections. 

The paper is organized as follows. At first we will solve the Cosmological system using our assumption and obtain the solutions analytically. Thereafter, we will build the major Cosmological equations using the new parametrization. Finally we will use observational data to constrain the parameter values and eventually use the observational results to explore the physical features of the model. 

\section{Mathematical Formulation}

We will assume that the Universe at Cosmological scale is spatially flat, homogeneous and isotropic. Then the geometry of the Universe at background level will satisfy the standard FLRW metric (in relativistic units $c = 1$) given by,
\begin{equation}
ds^2 = -dt^2 -a^{2}(t)\left[dr^2 + r^2(d{\theta}^2 + \sin^2 \theta d{\phi}^2)\right],
\label{metric}
\end{equation}
where $a(t)$ is the scale factor. Considering the components of Cosmological fluid are Baryons, radiation, Dark matter and Dark Energy, the Friedmann equation can be written as,

\begin{equation}
H^2 = \frac{8\pi G}{3}(\rho_b + \rho_{dm} + \rho_r + \rho_{de}) \label{FE}
\end{equation}
where $H = \frac{\dot{a}}{a}$ is the Hubble parameter and $G$ is Newton's gravitational constant. We will further assume that there is an interaction between the components of dark sector i.e. dark matter and dark energy are non-minimally coupled to each other. Additionally, Baryons and radiation will be considered to be conserved separately. Then, we obtain the following set of equations for Energy-momentum conservation of the components of cosmological fluid,    
\begin{equation}
\dot{\rho_{dm}} + 3H(1+\omega_{dm}){\rho_{dm}} = Q \label{conservation_dm}
\end{equation}

\begin{equation}
\dot{\rho_{de}} + 3H(1+\omega_{de}){\rho_{de}} = -Q \label{conservation_de}
\end{equation}

\begin{equation}
\dot{\rho_r} + 3H(1+\omega_r){\rho_r} = 0 \label{conservation_r}
\end{equation}

\begin{equation}
\dot{\rho_b} + 3H(1+\omega_b){\rho_b} = 0 \label{conservation_b}
\end{equation}
where $Q$ denotes the unknown interaction between the components of dark sector whereas $\omega$ denotes the equation of state parameter of the respective components. 
In accordance with our hypothesis, outlined in Introduction section, that in IDE models, Dark matter will follow a dilution law different from standard Cosmology, we write down the modified dilution rate of Dark matter density Energy as,

\begin{equation}
\rho_{dm} = \rho_{dm0}a^{-I}\label{dilution_law}
\end{equation}
where the dilution parameter\footnote{the concept of dilution parameter has been introduced in \cite{Aich} for $\Lambda$(t)CDM Cosmology. We have adopted the definition in the context of IDE model} in terms of scale factor and redshift respectively ($I$) in general is different from the standard expression 3(1+$\omega_{dm}$). Radiation and Baryons will follow their standard dilution law since they are self-conserved. The modification of dilution rate of Dark matter as a consequence of the interaction in dark sector is a very natural assumption that we will adopt in this work and use it to solve the system of equations (\ref{conservation_dm} -\ref{conservation_de}). Furthermore, since exact nature of interaction is unknown, we will adopt a specific expressions for the the unknown interaction term in a phenomenological manner - $Q$ = $\kappa(\dot\rho_{dm} +\dot\rho_{de})$

The conservation equation of dark matter (\ref{conservation_dm}) and dark energy (\ref{conservation_de}) written in terms of scale factor takes the form, 

\begin{equation}
\frac{d\rho_{dm}}{da} + \frac{3}{a}(1+\omega_{dm})\rho_{dm} = \kappa\left(\frac{d \rho_{dm}}{da} + \frac{d \rho_{de}}{da}\right)\label{conservation_dm_scale}
\end{equation}

\begin{equation}
\frac{d\rho_{de}}{da} + \frac{3}{a}(1+\omega_{de})\rho_{de} = - \kappa\left(\frac{d \rho_{dm}}{da} + \frac{d \rho_{de}}{da}\right)\label{conservation_de_scale}
\end{equation}

In standard Cosmology, Dark matter is usually assumed to be pressureless dust with E.O.S. $\omega_{dm}$ = 0. Studies with variable E.O.S. of dark matter have found negligible deviation from $\omega_{dm}$ = 0 \cite{Yang3}. Henceforth, in line with general consensus, we will also assume Dark matter to be pressureless dust in this work. Setting $\omega_{dm}$ = 0 and using (\ref{dilution_law}) in (\ref{conservation_dm_scale}) we obtain

\begin{equation}
\rho_{de} = \frac{\rho_{dm0}}{\kappa}\left[1-\frac{3}{I}-\kappa\right]a^{-I} + \tilde{\rho}_{de}\label{de_density}
\end{equation}

where $\tilde{\rho}_{de}$ is a constant of integration which denotes the value of dark energy density in the limit $a\rightarrow \infty$. 

Substituting (\ref{de_density}) in (\ref{conservation_de_scale}), we have an expression for E.O.S. OF dark energy,

\begin{equation}
\omega_{de} = -1 + \left[\frac{\frac{\rho_{dm0}}{\kappa}\left( 1 -\frac{3}{I}-\frac{3\kappa}{I}\right)a^{-I}}{\frac{\rho_{dm0}}{\kappa}\left(1-\frac{3}{I} - \kappa\right)a^{-I} + \tilde{\rho_{de}}}\right]\frac{I}{3}
\label{omega_de}
\end{equation}

\section{IDE Cosmology}
In the previous section, by hypothesizing a modified dilution law for evolution of dark matter, we obtained the expression of dark energy density and dark energy E.O.S. As a consequence, the standard equations of Cosmology will be modified. In this section we will evaluate the modified Cosmological expressions. The standard definition of dimensionless density parameters ($\Omega_i$) for $i^{th}$ component of cosmological fluid is,

\begin{equation}
 \Omega_{i} = \frac{8\pi G}{3H^2}\rho_i   
\label{dimensionless_density}
\end{equation}
Using (\ref{dimensionless_density}) The expressions of dimensionless dark matter and dark energy density can be written as, 
\begin{equation}
\Omega_{dm} = \frac{8\pi G}{3H^2}\rho_{dm0}a^{-I}\label{dimensionless_density_dm}
\end{equation}

\begin{equation}
\Omega_{de} = \frac{8\pi G}{3H^2}\left[\frac{\rho_{dm0}}{\kappa}\left(1-\frac{3}{I}-\kappa\right)a^{-I} + \tilde{\rho}_{de}\right]\label{dimensionless_density_de}
\end{equation}
Equation \ref{dimensionless_density_de} can be written as,
\begin{equation}
\Omega_{de} =  \Omega_{dm}\left(\frac{1}{\kappa}-\frac{3}{I\kappa}-1\right) + \tilde{\Omega}_{de} 
\label{dimensionless_density_de1}
\end{equation}
where we have defined $\tilde{\Omega}_{de} = \frac{8\pi G}{3H^2}\tilde{\rho}_{de}$. The expressions of dimensionless radiation and baryon density will be the usual standard expressions since these components of cosmological fluid are self-conserved. 

\subsection{Friedmann equation}
Using the definitions of dimensionless density parameters, friedmann equation can be written in terms of scale factor as,
\begin{equation}
H^2(a) = H_0^2\left[\Omega_{b0}a^{-3} + \Omega_{r0}a^{-4}+\Omega_{dm0}a^{-I}\left(1-\frac{3}{I}\right)\frac{1}{\kappa} + \tilde{\Omega}_{de0}\right]\label{friedmann_dimensionless}
\end{equation}
where the suffix zero represents present-day values of the density parameters following standard notation convention. Since the Universe has been assumed to be spatially flat, the dimensionless density parameters will satisfy the criteria,
\begin{equation}
\Omega_{b} + \Omega_{r} + \Omega_{dm} +\Omega_{de} = 1
\label{flat universe}
\end{equation}
Using \ref{flat universe}, we have,
\begin{equation}
\tilde{\Omega}_{de0} = 1 - \Omega_{b0} -\Omega_{r0} - \Omega_{dm0}\left(1-\frac{3}{I}\right)\frac{1}{\kappa}
\label{constant_de_density}
\end{equation}
Substituting \ref{constant_de_density} in \ref{friedmann_dimensionless}, we get,

\begin{multline}
H^2(z) = H_0^2\left[\Omega_{b0}(1+z)^{3} + \Omega_{r0}(1+z)^{4}+\Omega_{dm0}(1+z)^{I}\left(1-\frac{3}{I}\right)\frac{1}{\kappa}\right.\\
+ \left.1 - \Omega_{b0} -\Omega_{r0} - \Omega_{dm0}\left(1-\frac{3}{I}\right)\frac{1}{\kappa}\right]
\label{friedmann_dimensionless_redshift}
\end{multline}
where we have replaced scale factor by redshift parameter ($z$) by using the relation $a = \frac{1}{1+z}$. Equation \ref{friedmann_dimensionless_redshift} is the modified Friedmann equation for our model in terms of redshift parameter.

\subsection{Deceleration parameter}
An important parameter in Cosmology is the deceleration parameter ($q$) which measures whether the expansion of the Universe is accelerating or decelerating. It is given by,
\begin{equation}
q = - 1 - \frac{\dot{H}}{H^2} = -1 - \frac{\dot{a}}{H^2}\frac{dH}{da}
\label{dec.defn}
\end{equation}

Using Friedmann equation (\ref{friedmann_dimensionless}), an expression for scale factor as a function of cosmic time can be obtained as,

\begin{equation}
{\dot{a}}^2 = {H_{0}}^2 \left[ \Omega_{bo}a^{-1} + \Omega_{ro}a^{-2} + \Omega_{dmo}a^{2-I}\left(1- \frac{3}{I}\right)\frac{1}{\kappa} + \tilde{\Omega}_{deo}a^2\right] 
\label{scale_factor}
\end{equation}

Differentiating Friedmann equation (\ref{friedmann_dimensionless}) w.r.t. scale factor, we have,

\begin{multline}
\frac{dH}{da} = \frac{H_0}{2}\left[\Omega_{b0}a^{-3} + \Omega_{r0}a^{-4}+\Omega_{dm0}a^{-I}\left(1-\frac{3}{I}\right)\frac{1}{\kappa} + \tilde{\Omega}_{de0}\right]^{-1/2}\\  \times \left[(-3)\Omega_{b0}a^{-4} + (-4)\Omega_{r0}a^{-5}+ (-I)\Omega_{dm0}a^{-I-1}\left(1-\frac{3}{I}\right)\frac{1}{\kappa}\right]
\label{dH/da}
\end{multline}

Substituting (\ref{friedmann_dimensionless}), (\ref{scale_factor}) and (\ref{dH/da})  in (\ref{dec.defn}) we get,

\begin{multline}
q(a) = -1 - \left(\rule{0cm}{1cm}\right.\frac{{H_{0}} \left[ \Omega_{bo}a^{-1} + \Omega_{ro}a^{-2} + \Omega_{dmo}a^{2-I}\left(1- \frac{3}{I}\right)\frac{1}{\kappa} + \tilde{\Omega}_{deo}a^2\right]^{1/2}}{H_0^2\left[\Omega_{b0}a^{-3} + \Omega_{r0}a^{-4}+\Omega_{dm0}a^{-I}\left(1-\frac{3}{I}\right)\frac{1}{\kappa} + \tilde{\Omega}_{de0}\right]}\\  \times \frac{H_0}{2}\left[\Omega_{b0}a^{-3} + \Omega_{r0}a^{-4}+\Omega_{dm0}a^{-I}\left(1-\frac{3}{I}\right)\frac{1}{\kappa} + \tilde{\Omega}_{de0}\right]^{-1/2}\\  \times \left[(-3)\Omega_{b0}a^{-4} + (-4)\Omega_{r0}a^{-5}+ (-I)\Omega_{dm0}a^{-I-1}\left(1-\frac{3}{I}\right)\frac{1}{\kappa}\right]\left.\rule{0cm}{1cm}\right)
\label{dec.param}
\end{multline}

In terms of redshift, the expression of deceleration parameter can be written in a more convenient form as,
\begin{multline}
q(z) = -1 - \left(\rule{0cm}{1cm}\right.\frac{{H_{0}} \left[ \Omega_{bo}(1+z)^{1} + \Omega_{ro}(1+z)^{2} + \Omega_{dmo}(1+z)^{I-2}\left(1- \frac{3}{I}\right)\frac{1}{\kappa} + \tilde{\Omega}_{deo}(1+z)^{-2}\right]^{1/2}}{H_0^2\left[\Omega_{b0}(1+z)^{3} + \Omega_{r0}(1+z)^{4}+\Omega_{dm0}(1+z)^{I}\left(1-\frac{3}{I}\right)\frac{1}{\kappa} + \tilde{\Omega}_{de0}\right]}\\  \times \frac{H_0}{2}\left[\Omega_{b0}(1+z)^{3} + \Omega_{r0}(1+z)^{4}+\Omega_{dm0}(1+z)^{I}\left(1-\frac{3}{I}\right)\frac{1}{\kappa} + \tilde{\Omega}_{de0}\right]^{-1/2}\\  \times \left[(-3)\Omega_{b0}(1+z)^{4} + (-4)\Omega_{r0}(1+z)^{5}+ (-I)\Omega_{dm0}(1+z)^{I+1}\left(1-\frac{3}{I}\right)\frac{1}{\kappa}\right]\left.\rule{0cm}{1cm}\right)
\label{dec.param.redshift}
\end{multline}
Equation \ref{dec.param} and \ref{dec.param.redshift} gives the modified expression of deceleration parameter for our model in terms of scale factor and redshift respectively. In section 5, we will plot the evolution of deceleration parameter for IDE model. 

\subsection{Distances relations}
In Cosmology distances are measured in terms of luminosity of astrophysical objects (Luminosity distance) and angular diameter of astrophysical objects (angular diameter distance). For a flat Universe, they are defined as,

\begin{subequations}
\begin{equation}
d_{L}= \chi  \left( 1+z \right)
\end{equation}
\begin{equation}
d_{A}=\frac{ \chi }{ \left( 1+z \right) }
\end{equation}
\end{subequations}
where \(  \chi = \int _{0}^{z}\frac{dz}{H \left( z \right) } \). For the IDE model, the modified expressions of luminosity distance and angular diameter distance in terms of redshift will take the form,

\begin{multline}
d_L = \frac{1+z}{H_0}\int_{0}^{z}\frac{dz}{\Bigg[\Omega_{b0}(1+z)^{3} + \Omega_{r0}(1+z)^{4}+\Omega_{dm0}(1+z)^{I}\left(1-\frac{3}{I}\right)\frac{1}{\kappa}\Bigg.} \\
\Bigg. + 1 - \Omega_{b0} -\Omega_{r0} - \Omega_{dm0}\left(1-\frac{3}{I}\right)\frac{1}{\kappa}\Bigg]^{1/2}
\label{LD}
\end{multline}

\begin{multline}
d_A = \frac{(1+z)^{-1}}{H_0}\int_{0}^{z}\frac{dz}{\Bigg[\Omega_{b0}(1+z)^{3} + \Omega_{r0}(1+z)^{4}+\Omega_{dm0}(1+z)^{I}\left(1-\frac{3}{I}\right)\frac{1}{\kappa}\Bigg.} \\
\Bigg. + 1 - \Omega_{b0} -\Omega_{r0} - \Omega_{dm0}\left(1-\frac{3}{I}\right)\frac{1}{\kappa}\Bigg]^{1/2}
\label{AD}
\end{multline}

In particular, we will use the expression of Luminosity distance for our observational analysis in section 4, where the expression will build up in the expression of distance modulus which will be utilized in fitting type 1a supernova data against the model.

\subsection{Look back time and Cosmic Age}
Another important quantity in Cosmology is Look-back time, which is defined as the difference between the cosmic time in which a galaxy emitted a photon (t) and present cosmic time when it is received by us ($t_0$). Mathematically, it is defined as,
\begin{equation}
   t_0 - t =  \int_{0}^{z}\frac{dz}{(1+z)H(z)}
 \label{LBdef}
\end{equation}
For IDE model, look-back time will be given by,
\begin{multline}
t_0 - t = \int_{0}^{z}\frac{dz}{(1+z)H_0\Bigg[\Omega_{b0}(1+z)^{3} + \Omega_{r0}(1+z)^{4}+\Omega_{dm0}(1+z)^{I}\left(1-\frac{3}{I}\right)\frac{1}{\kappa}\Bigg.}\\
+ \Bigg.1 - \Omega_{b0} -\Omega_{r0} - \Omega_{dm0}\left(1-\frac{3}{I}\right)\frac{1}{\kappa}\Bigg]^{1/2}
\label{LBtime}
\end{multline}
Age of the Universe which refers to the difference of cosmic time between Big Bang (t = 0) and present epoch ($t_0$) for IDE model can be written from the expression of \ref{LBtime} as,
\begin{multline}
t_0 = \int_{0}^{\infty}\frac{dz}{(1+z)H_0\Bigg[\Omega_{b0}(1+z)^{3} + \Omega_{r0}(1+z)^{4}+\Omega_{dm0}(1+z)^{I}\left(1-\frac{3}{I}\right)\frac{1}{\kappa}\Bigg.}\\
+ \Bigg.1 - \Omega_{b0} -\Omega_{r0} - \Omega_{dm0}\left(1-\frac{3}{I}\right)\frac{1}{\kappa}\Bigg]^{1/2}
\label{age}
\end{multline}
Cosmic age is a very crucial parameter in Cosmology and in section 5, we will estimate its value for the IDE model.

\section{Observational Constraints}
After developing the basic parametrisation of the model in previous section, in this section, we will put some observational  constraints on the parameter values. For the purpose, we will perform a joint statistical analysis using type 1a supernova data and Hubble data. 

\subsection{Type 1a supernova}
In type 1a supernova method of Cosmological parameter analysis, type 1a supernovae are considered to be “standard candles” with fixed intrinsic luminosity. The model parameters are then extracted using the Luminosity Distance relation (\ref{LD}). Suppose $M$ is the absolute magnitude and $m$ is the apparent magnitude of type 1a supernova.Then the theoretical relation between the luminosity distance and apparent magnitude can be written as,
\begin{equation}
    m_{model} = 5log_{10}\left(\frac{d_L}{1 Mpc}\right) + 25 + M
\label{app.mag}
\end{equation}
In this work, we have used 1048 supernova from Pantheon \cite{Scolnic} for analysis. Pantheon supernova data is presented as $(m_{obs},z)$ pairs. The measured apparent magnitude ($m_{obs}$) of supernova  usually remain bugged by several nuisance parameters which are associated with stretch factor, colour, corrections from distance biases etc. Absolute magnitude ($M$) of a supernova also acts like a nuisance parameter. The apparent magnitude reported in Pantheon database is the corrected apparent magnitude obtained after determining and adjusting all the nuisance parameters, except absolute Magnitude, by BEAMS with Bias Corrections (BBC) method \cite{Kessler}. So, while using the Pantheon dataset, absolute magnitude ($M$) must also be fitted along with Cosmological parameters. We will not use present day Baryon density as a free parameter\footnote{In the model Baryon is not interacting with dark sector and is independently conserved. Therefore, we assume that its value will not change significantly from its value in $\Lambda$CDM model} and will set its value as $\Omega_{bo}H^2 = 224$ following Planck results \cite{Aghanim}. Besides, value of present day radiation density being very small, has been neglected in the fitting process\footnote{From this point on wards, radiation will be neglected in this manuscript} following convention. The fitted parameters for the analysis are  $H_0, I, \Omega_{dm0}f(I, \kappa)$ and $M$ where $f(I,\kappa) = \left(1-\frac{3}{I}\right)\frac{1}{\kappa}$. The goodness of fit parameter for supernova analysis can then be defined as,

\begin{equation}
    \chi^2_{SN} = \frac{\left(m_{obs_i} - m_{model}\right)^2}{\sigma_{i}^2}
    \label{chisq.sn}
\end{equation}

Here $m_{obs_i}$ and $\sigma_{i}$ denotes the  observational value of apparent magnitude and uncertainty corresponding to the redshift $z_i$. $m_{model}$ is model dependent theoretical value of apparent magnitude obtained from (\ref{app.mag}).

\subsection{H(z) method}
In H(z) method, Hubble data obtained by Angular Diameter method and/or BAO method, is directly implemented to extract Cosmological parameters by utilising Friedmann equation \ref{friedmann_dimensionless_redshift}. For this work, we have used the Hubble data from table II of \cite{Sharov2}. The fitted parameters for the analysis are same as in supernova method with the exception of absolute magnitude $M$. 
The goodness of fit parameter for H(z) method can then be defined as
\begin{equation}
    \chi^2_{H(z)} = \frac{\left(H(z)_{obs_i} - H(z)_{model}\right)^2}{\sigma_{i}^2}
    \label{chisq.H(z)}
\end{equation}

Here $H(z)_{obs_i}$ and $\sigma_{i}$ denotes the  observational value of Hubble parameter and uncertainty corresponding to the redshift $z_i$. $H(z)_{model}$ is model dependent theoretical value of Hubble parameter obtained from \ref{friedmann_dimensionless_redshift}.

\subsection{Joint analysis and parameter estimates}
In this work, we have performed a joint analysis combining both supernova and Hubble data as described above. The goodness of fit parameter for the combined analysis is given by,
\begin{equation}
      \chi^2 =  \chi^2_{SN} + \chi^2_{H(z)}
      \label{chisq}
\end{equation}

The fitting has been carried out using non-linear least square method which has been implemented in Lmfit module \cite{Newville} of python. The parameter estimates has been shown in table \ref{results}. Figures (\ref{fig:confidence1}-\ref{fig:confidence6}) shows the parameter confidence intervals from the joint analysis.

\begin{figure}
\centering
\begin{subfigure}[b]{1\textwidth}
\centering
\includegraphics[width=0.9\textwidth]{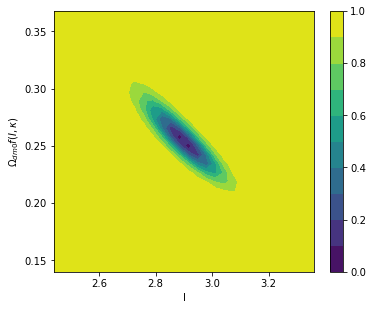}
\caption{Confidence interval of $I$ and $\Omega_{dm0}f(I, \kappa)$}
\label{fig:confidence1}
\end{subfigure}
\begin{subfigure}[b]{1\textwidth}
\centering
\includegraphics[width=0.9\textwidth]{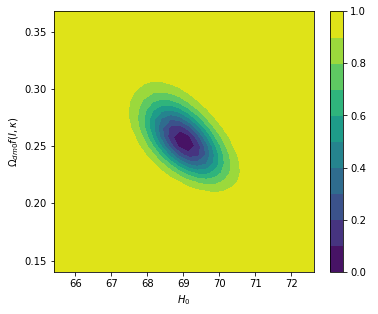}
\caption{Confidence interval of $H_0$ and $\Omega_{dm0}f(I, \kappa)$}
\label{fig:confidence2}
\end{subfigure}
\end{figure}

\begin{figure}
\ContinuedFloat
\centering
\begin{subfigure}[b]{1\textwidth}
\centering
\includegraphics[width=0.9\textwidth]{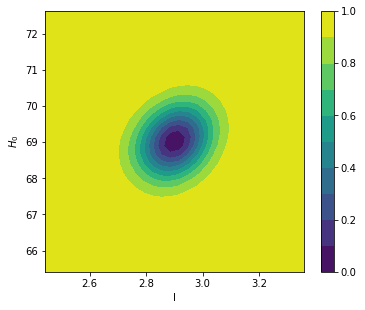}
\caption{Confidence interval of $I$ and $H_0$}
\label{fig:confidence3}
\end{subfigure}
\begin{subfigure}[b]{1\textwidth}
\centering
\includegraphics[width=0.9\textwidth]{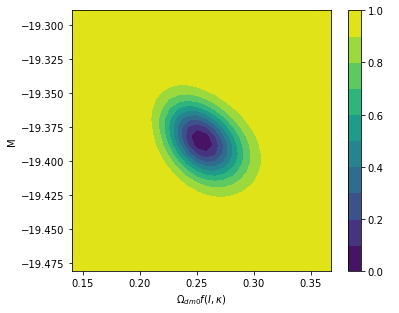}
\caption{Confidence interval of $M$ and $\Omega_{dm0}f(I, \kappa)$}
\label{fig:confidence4}
\end{subfigure}
\end{figure}

\begin{figure}
\ContinuedFloat
\centering
\begin{subfigure}[b]{1\textwidth}
\centering
\includegraphics[width=0.9\textwidth]{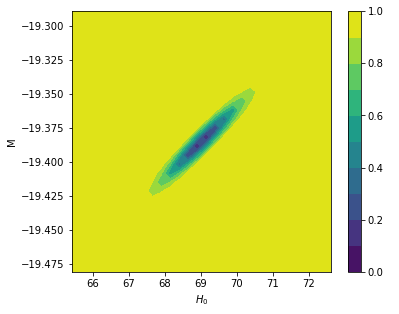}
\caption{Confidence interval of $H_0$ and $M$}
\label{fig:confidence5}
\end{subfigure}
\begin{subfigure}[b]{1\textwidth}
\centering
\includegraphics[width=0.9\textwidth]{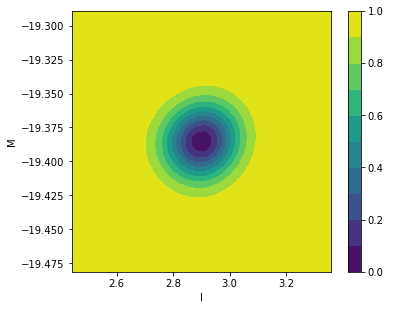}
\caption{Confidence interval of $I$ and $M$}
\label{fig:confidence6}
\end{subfigure}
\end{figure}

\begin{table}
    \centering
    \begin{tabular}{|c|c|}
    \hline
         \textbf{Cosmological Parameters} & \textbf{Values \ }\\
    \hline
        $\Omega_{dm0}f(I, \kappa)$ & 0.254 $\pm$ 0.023\\
    \hline
         $I$ & 2.901 $\pm$ 0.092 \\
    \hline
         $H_0$ &  69.023 $\pm$ 0.722 \\
    \hline
        $M$ &  -19.385 $\pm$ 0.019 \\
    \hline
    \end{tabular}
    \caption{fit results}
    \label{results}
\end{table}

\section{Physical features of the model}
In this section, we will study the physical features of the model using the observational constraints obtained in previous section. 
\subsection{Evolution of Universe}
The evolution of Universe in Cosmology is conveniently described by the evolution of scale factor as a function of cosmic time.
In terms of dimensionless relative time parameter, defined by $\tilde{t} = H_0(t - t_0)$, equation \ref{scale_factor} (with radiation neglected) can be written as,
\begin{equation}
  \left(\frac{da}{d\tilde{t}}\right)^2 = \left[ \Omega_{bo}a^{-1} + \Omega_{dmo}a^{2-I}\left(1- \frac{3}{I}\right)\frac{1}{\kappa} + \tilde{\Omega}_{deo}a^2\right]
\end{equation}

\begin{figure}
\centering
\includegraphics[width=1\textwidth]{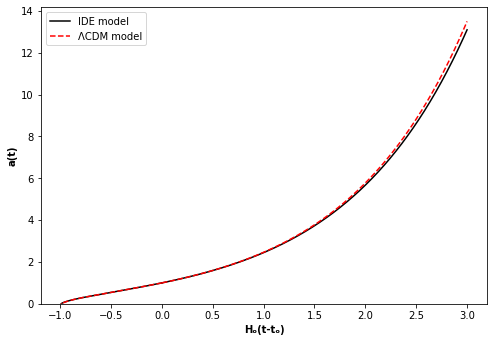}
\caption{Scale factor evolution as a function of redshift for IDE model and $\Lambda$CDM model}
\label{fig:scale_factor}
\end{figure}

The plot of scale factor as a function of dimensionless relative time parameter is shown in figure \ref{fig:scale_factor}. For comparison, we have also plotted the scale factor curve of $\Lambda$CDM model. The plots show that the evolution of $\Lambda$CDM Universe and IDE universe is similar. At very late universe, there is a visible splitting of the two curves and there is indication that IDE Universe evolves marginally slower compared to compared to $\Lambda$CDM universe. 

\subsection{Deceleration parameter evolution and transition redshift}
\begin{figure}
\centering
\includegraphics[width=1\textwidth]{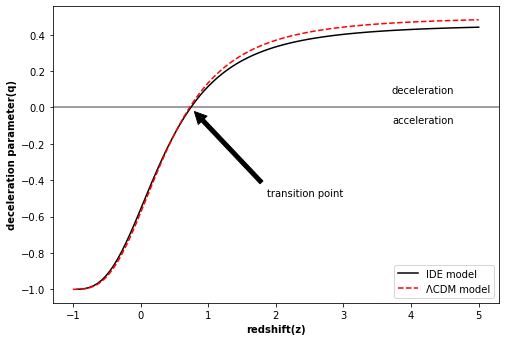}
\caption{Deceleration parameter evolution as a function of redshift for IDE model and $\Lambda$CDM model}
\label{fig:dec_param}
\end{figure}

Using the observational constraints of previous section, a plot of deceleration parameter as a function of redshift can be obtained from equation (\ref{dec.param.redshift}) which is shown in figure \ref{fig:dec_param}. Deceleration parameter curve for $\Lambda$CDM model has also been plotted for comparison.

The present day value of deceleration parameter and the transition redshift corresponding to the null value of deceleration parameter for the model can be estimated from figure \ref{fig:dec_param} as, 

\begin{center}
$q_0^{IDE} = -0.561$ ; $z_t^{IDE} =  0.744$
\end{center}
The values marginally deviates from $\Lambda$CDM model which yields values,
\begin{center}
$q_0^{\Lambda CDM} = -0.580$ ; $z_t^{\Lambda CDM} =  0.727$
\end{center}
The slight difference of the parameter values from $\Lambda$CDM counterparts is a reflection of the slower evolution of IDE Universe as seen in figure (\ref{fig:scale_factor})

\subsection{Cosmic Age}
The Cosmic Age for IDE model can be estimated from \ref{age} using the best-fit values obtained in previous section. It gives,
\begin{center}
$t_0^{IDE} = 14.032$ Gyr
\end{center}
Evidently, the value of Cosmic Age is slightly higher than $\Lambda$CDM model which yields, 
\begin{center}
$t_0^{\Lambda CDM} = 13.885$ Gyr
\end{center}
The higher value of $t_0$ in IDE model in comparison to $\Lambda$CDM model corresponds to the slower evolution of the Universe as seen in figure (\ref{fig:scale_factor}).

\subsection{Particle Physics implications}
The estimated value of dilution parameter is less than the standard value i.e. 3 which is a consequence of the interaction between the two components of dark sector. Even though the exact nature of the interaction mechanism is unknown, we can still study some of the consequences from the estimated value. In regards to the lower value of dilution parameter than standard case, two possible physical scenarios can emerge - (i) extra dark matter particles are being created as a consequence of the interaction which results in a slower dilution of Dark matter compared to standard model. (ii) the mass of Dark matter particles are varying due to the interaction i.e. Dark matter particles are effectively "Variable Mass Particles" (VAMP).  

If $m_{dm}$ and $n_{dm}$ be the mass and number density of dark matter, then energy conservation equation of dark matter (\ref{conservation_dm_scale}) can be written as,
\begin{equation}
\frac{d\left(m_{dm}n_{dm}\right)}{da} + \frac{3}{a}(1+\omega_{dm})m_{dm}n_{dm} = \kappa\left[\frac{d \left(m_{dm} n_{dm}\right)}{da} + \frac{d \rho_{de}}{da}\right]
\label{conservation_dm_mn}
\end{equation}
\subsubsection{Case I: Dark matter creation}
Assuming the Dark matter particles are mass invariant and setting $\omega_{dm} = 0$, (\ref{conservation_dm_mn}) can be written as,
\begin{equation}
    m_{dm}\frac{dn_{dm}}{da} + \frac{3}{a}m_{dm}n_{dm} = \kappa \left(m_{dm}\frac{dn_{dm}}{da} + \frac{d \rho_{de}}{da}\right)
\label{conservation_dm_mn_1}
\end{equation}
Equation \ref{conservation_dm_mn_1} can be written in a convenient form,
\begin{equation}
  \frac{dn_{dm}}{da} + \frac{3}{a\left(1-\kappa\right)}n_{dm} = \frac{\kappa}{1-\kappa} \frac{n_{dm}}{\rho_{dm}} \frac{d\rho_{de}}{da}
\label{conservation_dm_mn_1_convenient}
\end{equation}
Equation \ref{conservation_dm_mn_1_convenient} can solved to obtain,
\begin{equation}
    n_{dm} = n_{dm0} a^{-I}
\label{particle_density}
\end{equation}
where $n_{dm0}$ is the present day value of number density of dark matter. 
\subsubsection{Case II: VAMP scenario}
In this case, we will allow mass of Dark matter to vary while simultaneously requiring that there is no creation of Dark matter particles. For this case, equation \ref{conservation_dm_mn} takes the form,
\begin{equation}
m_{dm}\frac{dn_{dm}}{da} + n_{dm}\frac{dm_{dm}}{da} + \frac{3}{a}m_{dm}n_{dm} = \kappa\left[m_{dm}\frac{dn_{dm}}{da} + n_{dm}\frac{dm_{dm}}{da} + \frac{d \rho_{de}}{da}\right]
\label{conservation_dm_mn_2}
\end{equation}
Assuming no extra Dark matter particles are created, the dark matter particle number density must follow the standard equation,
\begin{equation}
    \frac{dn_{dm}}{da} + \frac{3}{a}n_{dm} = 0
\label{standard_no._density}
\end{equation}
The system of equations \ref{standard_no._density} and \ref{conservation_dm_mn_2} can be solved to obtain,
\begin{equation}
    m_{dm} = m_{dm0}a^{3-I}
\end{equation}
where $m_{dm0}$ is the present day value of Dark matter mass.

\section{Discussions and Conclusion}
In this work, we introduced a new approach to analyse Interacting Dark Energy models where instead of using the conventional dynamical system analysis, we hypothesised that in IDE scenario, Dark matter density will dilute in a rate different from standard non-interacting case. Based on this assumption, we obtained a new parametrization in terms of modified dilution rate of Dark matter, dubbed dilution parameter. The entire Cosmology was then built revolving around the dilution rate parametrization and the model was confronted against observational data which yielded the parameter estimates, $\Omega_{dm0}f(I,\kappa) = 0.254, I = 2.901, H_0 = 69.023, M = -19.385$. The derived parameters were estimated to be, $q_0 = - 0.561, z_t = 0.744, t_0 =14.032$ Gyr. The value of dilution parameter ($I$) obtained was less than the standard value of non-interacting scenario which indicates that Dark matter dilutes slower in IDE Universe. Physically, this can mean either extra Dark matter particles are being created or mass of dark matter is not constant and is increasing. In these possible physical situations, transfer of energy has to be from Dark Energy to Dark matter. The plots of evolution of scale factor and deceleration parameter showed similar pattern with $\Lambda$CDM model, but the IDE Universe seemed to evolve slower in comparison with $\Lambda$CDM Universe. The marginal deviation of the derived parameter values from $\Lambda$CDM model also seems to reflect the slower evolution. The plots further reflects that the model is very close to $\Lambda$CDM model which means that the interacting scenario cannot be distinguished from non-interacting scenario, given the current state of precision in observational methods. However, this also means that the interacting scenario cannot be ruled out at present and must be investigated further.

\end{document}